\documentstyle[here,12pt]{article}

\catcode`\@=11

\def\section{\@startsection{section}{1}{\z@}{3.5ex plus 1ex minus
   .2ex}{2.3ex plus .2ex}{\bf}}

\catcode`\@=12

\newfont{\mbm}{msbm10 scaled\magstep1}
\def\bb#1{\hbox{\mbm #1}}

\newcommand{\smallfrac}[2] {\mbox{$\frac{#1}{#2}$}}
\newcommand {\eqref} [1] {(\ref {#1})}
\newcommand {\beq} {\begin{equation}} 
\newcommand {\eeq} {\end{equation}}
\newcommand {\ber}{\begin{eqnarray*}}
\newcommand {\eer} {\end{eqnarray*}}
\newcommand {\bea}{\begin{eqnarray}}
\newcommand {\eea} {\end{eqnarray}}

\newcommand{\Nfour} {${\cal N}=4\ $}
\newcommand{\Ntwo}{${\cal N}=2\ $}

\def\drawbox#1#2{\hrule height#2pt
        \hbox{\vrule width#2pt height#1pt \kern#1pt
              \vrule width#2pt}
              \hrule height#2pt}

\def\Asym#1#2{\vcenter{\vbox{\drawbox{#1}{#2}
              \kern-#2pt       
              \drawbox{#1}{#2}}}}

\begin{document}\begin{titlepage}
\rightline{{CPTH-S004.0300}}
\rightline{{LPTENS 00/07}}
\rightline{{hep-th/0003050}}
\vskip 2cm
\centerline{{\large\bf RG Flow, Wilson Loops and the Dilaton Tadpole}}
\vskip 1cm
\centerline{{Carlo Angelantonj}${}^\dagger$ and 
{Adi Armoni}${}^\ddagger$}
\vskip 0.5cm
\centerline{\it ${}^\dagger$ Laboratoire de Physique Th{\'e}orique de
l'{\'E}cole Normale Sup{\'e}rieure\footnote{Unit{\'e} mixte du CNRS et de
l'ENS, UMR 8549}}
\centerline{\it 24, rue Lhomond, F-75231 Paris Cedex 05}
\centerline{\tt angelant@lpt.ens.fr}
\vskip 0.3cm 
\centerline{\it ${}^\ddagger$ Centre de Physique Th{\'e}orique de l'{\'E}cole 
Polytechnique\footnote{Unit{\'e} mixte du CNRS et de l'EP, UMR 7644}}
\centerline{\it F-91128 Palaiseau Cedex}
\centerline{\tt armoni@cpht.polytechnique.fr}
\vskip  1.0cm
\begin{abstract}
We discuss the role of the dilaton tadpole in the holographic description
of non-supersymmetric gauge theories that are conformal in the planar limit.
On the string theory side, the presence of the dilaton tadpole modifies  
the AdS background inducing a logarithmic 
running for the radius and the dilaton.
Using the holographic prescription we compute the Wilson loop on the
gravity side and find a smooth interpolating potential between 
asymptotic freedom and confinement, as expected from field theory.
\end{abstract}
\end{titlepage}

\section{Introduction}

The AdS/CFT correspondence has improved considerably our
understanding of strongly coupled large $N$ gauge theories \cite{ads},
showing that the natural description of 4d theories in 
this regime is in terms of
5d gravity, where the extra dimension is related to the
4d energy scale. So far, strong evidence for the validity of the 
conjecture has come from the description of \Nfour SYM in terms of IIB
supergravity, while limited efforts have been devoted to the 
holographic description of non-supersymmetric non-conformal field 
theories. Moreover, the few examples analyzed were mostly based on the 
tachyonic (oriented) 0B string \cite{KT1,BFL2}, where the
presence of the tachyon complicates the gravity description
\cite{minahan,log} and leads to instabilities in the 
dual strongly coupled field theory \cite{klebanov}. 

The first attempt to find a {\it non-tachyonic} holographic 
description of non-supersymmetric gauge theories was recently made
in \cite{AA}. The analysis was still performed within the 
framework of 0B strings, but resorting to a particular orientifold 
projection \cite{sagnotti2} to eliminate the tachyon from 
the bulk and to avoid problems with the doubling of R-R sectors. 
Actually, the gauge theories that naturally emerge from
(orientifolds of) type 0 models have the peculiar property of becoming
conformal in the planar limit, and therefore several
results can be mutuated from \Nfour \cite{KS,AA}. 
In particular, the leading (planar)
geometry is known to be of the form $AdS_5 \times X_5$ with a constant
dilaton field. For finite $N$, however, the gauge theory is no longer
conformal and new features are expected in the gravity description. 
Indeed, a general property of non-supersymmetric theories seems to be
the presence of a non-vanishing dilaton tadpole at genus one-half 
\cite{ADS}. On the other hand, supersymmetry relates R-R tadpoles to NS-NS
ones, and thus demanding neutral configurations of orientifold
planes and D-branes automatically ensures the vanishing of all
massless tadpoles. However, whenever supersymmetry is not present one
can find configurations for which the R-R and NS-NS tadpole conditions are
incompatible \cite{sagnotti2,ADS,BFL1}. While uncanceled R-R tadpoles
manifest themselves in pathologies of the string theory \cite{PC},
NS-NS tadpoles are less problematic. They correspond to potential
terms in the low-energy action and their main effect is to modify the
vacuum. In our case this is precisely what one needs! 
Indeed, studying D3 branes in
non-tachyonic orientifolds of type 0B (with ${\rm O}'7$-planes and
D7-branes, where a prime indicates the non-tachyonic involution) 
we showed explicitly \cite{AA} how a genus one-half dilaton tadpole 
in the bulk theory does reproduce qualitatively and quantitatively 
the expected logarithmic gauge theory RG flow. (Recently, 
a logarithmic running of the gauge coupling was also found in 
the gravitational description of \Ntwo ${\rm SU}(N)
\times {\rm SU} (N+M)$ gauge theory \cite{KNT}. See related 
discussion on dilaton-induced RG flow in \cite{AlGo}.)
 
The purpose of this letter is to pursue this program further,
studying the back-reaction on the metric induced by the presence
of a dilaton tadpole. 
In particular, we compute the 
Wilson loop in string theory and find a quark anti-quark potential
interpolating between a (logarithmically running) Coulomb phase
and a confining phase in the IR, as expected from the field theory 
analysis. To the best of our knowledge, the interpolating solution
that we find is the only known example in which a single
solution describes the expected behavior in the entire energy
regime. Although our analysis is quite general and applies to a vast  
class of non-supersymmetric non-conformal theories for which a 
non-vanishing dilaton tadpole is expected to develop \cite{ADS}, the
simplest model one may conceive originates from D3-branes and ${\rm
O}'3$-planes in type 0B. In the large $N$ limit the near horizon 
geometry is $AdS_5
\times \bb{RP}^5$ \cite{witten}, where $\bb{RP}^5 = S^5 /\bb{Z}_2$, 
with $\bb{Z}_2$ the
non-tachyonic involution \cite{sagnotti2} associated to ${\rm
O}'3$. Since the $\bb{Z}_2$ acts freely on $S^5$, no open sector is present
in the bulk and the spectrum of the model consists of all the 
$\bb{Z}_2$-invariant harmonics of the 0B fields. In particular, 
the tachyon is projected out. Moreover, since two-cycles of
$\bb{RP}^5$ are unorientable, as for instance $\bb{RP}^2 \subset \bb{RP}^5$,
new interactions in the low-energy action are expected to originate from
unoriented world-sheet topologies. These are weighted by generic
(not only even) powers of the dilaton, and in particular a dilaton
tadpole can be generated. Unfortunately, an explicit computation of
one-point amplitudes for the dilaton in the curved $AdS_5
\times \bb{RP}^5$ geometry seems out of reach within the actual
(perturbative) definition of string theory, although such a term is
expected from the above arguments, from our experience with
non-supersymmetric theories, and from the results we are going to
present here.

The organization of this letter is as follows: in section 2
we introduce the class of field theories we want to study, and
present an explicit example from (unoriented) D-branes in the 0B string.
In section 3 we turn to the gravity description and find an explicit 
solution for the dilaton and the metric up to next-to-leading order in
$1/N$ expansion.
In section 4 we extract the holographic quark anti-quark potential 
and compare it with the field theory analysis of section 2. 
Finally, section 5 contains our final comments about the validity of
the gravity description.

\section{Field theory considerations}

One of the most remarkable results in field theory is the running of
coupling constants
\beq
{d \alpha (u) \over d \log u}  = - \beta (\alpha )\ ,
\eeq
where $\alpha (u) = g^2_{{\rm YM}} (u) \, N$ is the 't Hooft coupling.
The precise expression of the beta-function is model dependent 
and encodes the quantum
corrections to the classical behavior $\alpha (u) = \alpha
_0$. Whenever a quantum theory does not depend on any scale and thus
is exactly conformally invariant, as for instance \Nfour SYM, $\beta (\alpha)$ 
vanishes identically and the coupling constants take their classical 
values for all energy scales. Combining these two arguments, it is then
evident that theories that are conformally invariant in the planar
limit are governed by a RG equation
\beq
{d \alpha (u) \over d \log u}  = -{1\over N}\, \beta _1 (\alpha )
- {1\over N^2}\, \beta _2 (\alpha) \, + ... \ , 
\label{beta}
\eeq
so that one recovers the expected 
$\alpha (u) = \alpha_0$ result for $N\to\infty$. Here, $\beta_i
(\alpha)$ {\it do not} represent the $i^{{\rm th}}$-loop contribution
to the RG equation. Rather, they encode {\it all} the contributions
at a given order in $1/N$.

To be concrete, the simplest prototype one can conceive is an
${\rm SU}(N)$ gauge theory with six real scalars in the adjoint
and 4 Weyl spinors in the antisymmetric and antisymmetric-conjugate 
representations, arising from the non-tachyonic orientifold projection 
on type 0B D3 branes \cite{sagnotti2,BFL1}. This theory was shown 
to converge to the ${\rm SU}(N)$ \Nfour SYM in the planar limit 
\cite{AA}, and to have a one-loop beta-function suppressed by 
$1/N$ (with $b_1 = {16\over 3}$), 
and a two-loop beta-function contributing both to $\beta_1 (\alpha)$
and $\beta_2 (\alpha )$ in \eqref{beta} \cite{BFL2}.

One can extract some interesting results from the RG equation
\eqref{beta}. Since in \eqref{beta} there is a natural small
parameter, $1/N$, it is natural to solve this equation by iterations
\beq
\alpha (u) = \alpha _0 + {1\over N} \alpha _1 + \ldots \ .
\eeq
The solution
\beq
\alpha (u) = \alpha _0 - {1\over N} \beta _1 (\alpha _0) \log u
\eeq
that, to this order, can be equivalently written as
\beq
{1\over \alpha(u)} = {1\over \alpha _0} + {1\over N} {\beta _1 (\alpha
_0) \over
\alpha_0 ^2} \log u \ ,\label{1loop}
\eeq
leads then to a quark anti-quark potential 
\beq
E(L) = {\alpha (L)\over L}  = 
{\alpha _0 \over L(1- {1\over N}{\beta _1 (\alpha_0)
 \over \alpha _0} \log L)} \ ,\label{asymptotic}
\eeq
as expected for asymptotically free theories. Actually, since for
large $N$ the beta-function varies slowly, the logarithmic behavior 
\eqref{1loop} persists beyond the perturbative regime, up to 
energy scales such that $\left| {1\over N} {\beta _1 (\alpha _0) \over 
\alpha _0 ^2} \log u \right| \ll 1$. Note that the expansion is not valid
in the (extreme) IR.

The description of the (non-perturbative) IR regime is less
straightforward. Although for the specific model we introduced previously the
scalars do not acquire a potential at the one-loop level \cite{BFL2,AA}, 
in general one would not expect to find a moduli space of vacua for 
non-supersymmetric theories. Thus, a potential for the
scalars should develop: the scalars might become massive, or tachyonic,
and might also acquire a non-vanishing vacuum expectation
value. Although the specific shape of the potential is model
dependent, in the following we shall assume that the scalars become 
massive without acquiring any {\it vev}. As we shall see, the string 
theory description supports this assumption. 

Once the massive scalars decouple, the IR degrees of freedom will
consist of Yang-Mills theory with four massless spinors 
(a QCD like theory), and a confining potential is expected
\beq
E(L) = \sigma  L \ , 
\eeq
where $\sigma = \Lambda _{{\rm QCD}} ^2$ is the QCD string tension. 
For pure (large $N$) Yang-Mills theory one expects 
a constant (finite) string tension $\sigma_0$. 
In the present case, however, one expects a different behavior. 
Since the theories we are describing are conformal in the planar 
limit, the string tension should to vanish accordingly. 
To estimate the $N$ dependence of the string tension, one can
integrate the RG equation \eqref{beta} from an arbitrary weak
coupling scale ($\sim u_0$ in the UV) to a strong coupling one 
($\sim \Lambda _{{\rm QCD}}$):
\beq
\Lambda _{{\rm QCD}} = u_0 \, e ^{-N \int {d\alpha \over \beta _1
(\alpha)}}\ ,
\eeq
that translates into an exponentially suppressed string tension
\beq
\sigma \sim e^{-N} \ . \label{stringtension}
\eeq

\section{The gravity solution}

We now turn to a dual gravitational description of the gauge
theories introduced in the previous section, in the spirit of the AdS/CFT
correspondence \cite{ads}. 
As discussed in \cite{AA}, the dual gravity description involves the
ten-dimensional metric, the dilaton and the R-R four-form potential (with
self-dual field strength), whose dynamics is essentially 
encoded in the action\footnote{To be more precise one should use the
PST action \cite{pst}, since we are dealing with a self-dual 5-form $F_5$.} 
\beq
S =
{1\over (\alpha ')^4} \int d^{10} x \sqrt {-g}
\Biggl[ e^{-2\Phi} (R+4 \partial _\rho \Phi \partial ^\rho \Phi ) + {1\over
\alpha '} C e^{-\Phi} -
\smallfrac{1}{4} (\alpha')^4  |F_5|^2 \Biggr] \ ,
\label{action} 
\eeq
or in its equations of motion
\bea
e ^{-2\Phi} \left ( 8 {(\nabla \Phi )}^2 - 8 \nabla ^2 \Phi -2R
 \right ) - Ce ^{-\Phi} &=& 0 \ ,
\label{motion1} 
\\
e ^{-2\Phi} \left( R_{\mu\nu} + 2 \nabla _\mu \nabla _\nu \Phi
 \right) + \smallfrac{1}{4} g_{\mu\nu} \, C\, e^{-\Phi} 
 \qquad \qquad & & \nonumber
\\ 
- \smallfrac{1}{96}
(F_{\mu \rho\sigma\lambda\tau} F_{\nu}{}^{\rho\sigma\lambda\tau} - 
\smallfrac{1}{10} g_{\mu\nu} 
F_{\rho\sigma\lambda\tau\eta} F^{\rho\sigma\lambda\tau\eta} ) 
&=& 0 \ . \label{motion2}
\eea
The term $C e^{-\Phi} = C\, g_{s}^{-1}$ corresponds to the
dilaton tadpole expected from half-genus in string perturbation theory. 
Although the direct computation of this term is difficult, 
there are plausible arguments in favor of its presence, as discussed in 
the introduction.

In the spirit of AdS/CFT correspondence, we make for the metric the ansatz
\beq
ds^2 = \alpha ' \left( d\tau ^2 + e^{2\lambda(\tau)} \eta^{ab} dx_a d x_b + 
e ^{2\nu (\tau)} d\Omega _5 ^2 \right) \ , \label{metric}
\eeq
compatible with four-dimensional Poincar\'e invariance and SU(4) 
flavor symmetry. Moreover, the five-form field strength has $N-{1\over
2}$ units of flux induced by the $N$ D3 branes and 
the O$'3$ plane. In the following we shall ignore the
contribution of the orientifold plane, manifestly suppressed in the 
presence of a large number of branes. 

Inserting the above ansatz in the eqs. (\ref{motion1}) and
(\ref{motion2}) and allowing for a $\tau$ dependence of the various fields
one obtains
\bea
2 \ddot \Phi -4\ddot \lambda -5 \ddot \nu - 4 \dot \lambda ^2 -5 \dot
\nu ^2 + \smallfrac{1}{2} N^2 e ^ {2\Phi - 10\nu} - \smallfrac{1}{4}
 C e ^\Phi &=& 0 \ ,
\label{eq1}
\\
\ddot \lambda - (2 \dot \Phi -4 \dot \lambda -5 \dot \nu) \dot \lambda
- \smallfrac{1}{2} N^2 e ^{2 \Phi -10\nu} +\smallfrac{1}{4} C e ^\Phi
&=& 0 \ ,
\label{eq2} 
\\
\ddot \nu - (2 \dot \Phi -4 \dot \lambda -5 \dot \nu) \dot \nu 
- 4 e ^ {-2\nu} + \smallfrac{1}{2} N^2 e ^ {2 \Phi - 10 \nu}  
+\smallfrac{1}{4}C e ^\Phi &=& 0 \ ,
\label{eq3} 
\\
-12 \dot \lambda ^2 - 20 \dot \nu ^2 -4 \dot \Phi ^2 + 20 \dot \Phi
\dot \nu +16 \dot \Phi \dot \lambda - 40 \dot \lambda \dot \nu & &
\nonumber
\\
+ 20 e ^ {-2\nu} - N^2 e ^{2\Phi -10\nu} + C e ^\Phi &=& 0 \ . \label{eq4}
\eea
As usual in General Relativity, the $\tau \tau$ component (\ref{eq4}) 
of Einstein's equations is not independent, and translates into a vanishing
energy constraint for the associated mechanical system.
  
In order to better appreciate the role of the different contributions
in the low-energy action, it is useful to redefine 
$\Phi \rightarrow \Phi - \log N$ , in order that $e ^\Phi$ 
be directly related to the 't Hooft coupling. The
(independent) field equations then read
\bea
2 \ddot \Phi -4 \dot \Phi ^2 + 8 \dot \Phi \dot \lambda
+10 \dot \Phi \dot \nu  +3 {C\over N} e ^\Phi &=& 0 \ ,
\label{phieq}
\\
\ddot \lambda - (2 \dot \Phi -4 \dot \lambda -5 \dot \nu) \dot \lambda
- \smallfrac{1}{2} e ^{2 \Phi -10\nu} +\smallfrac{1}{4} {C\over N} e
^\Phi &=& 0 \ ,
\label{lameq}
\\
\ddot \nu - (2 \dot \Phi -4 \dot \lambda -5 \dot \nu) \dot \nu 
- 4 e ^ {-2\nu} + \smallfrac{1}{2} e ^ {2 \Phi - 10 \nu}  
+\smallfrac{1}{4} {C\over N} e ^\Phi &=& 0 \ .
\label{nueq} 
\eea
From these one can neatly understand the role of the dilaton
tadpole: it represents a $1/N$ correction to the leading type IIB 
supergravity equations, and thus induces $1/N$
corrections to the $AdS_5 \times \bb{RP}^5$ leading geometry, in agreement
with field theory expectations. Moreover, $1/N$ is a natural
small parameter in the theory and suggests to solve the equations 
by iterations
\bea
\Phi (\tau ) &=& \Phi _0 (\tau) + {1\over N} \Phi _1 (\tau) + \ldots \ ,
\\
\lambda (\tau ) &=& \lambda _0 (\tau) + {1\over N} \lambda _1 (\tau) + 
\ldots \ , 
\\
\nu (\tau ) &=& \nu _0 (\tau) + {1\over N} \nu _1 (\tau) + \ldots 
\ . 
\eea

The leading order equations
\bea
2 \ddot \Phi _0 -4  \dot \Phi _0 ^2  + 8 \dot
\Phi _0 \dot \lambda _0
+10 \dot \Phi _0 \dot \nu _0 &=& 0 \ ,
 \\
\ddot \lambda _0 - (2 \dot \Phi _0 -4 \dot \lambda _0
-5 \dot \nu _0 ) \dot \lambda _0
- \smallfrac{1}{2} e ^{2 \Phi _0 - 10\nu _0} &=& 0 \ , 
 \\
\ddot \nu _0 - (2 \dot \Phi _0 -4 \dot \lambda _0 -5
\dot \nu _0 ) \dot \nu _0 
- 4 e ^ {-2\nu _0 } + \smallfrac{1}{2} e ^ {2 \Phi _0 -
 10 \nu _0 } &=& 0 \  ,
\eea
independent of the dilaton tadpole, are consistently solved by the 
expected (leading) $AdS_5 \times \bb{RP}^5$ geometry
\bea 
& & \Phi _0 = \varphi \ ,
\\
& & \lambda _0 = {\root 8 \of 8}\, e ^{-\smallfrac{1}{4} 
   \varphi} \tau \ ,
\\
& & \nu _0 = - \smallfrac{1}{8} \log 8 + \smallfrac{1}{4} \varphi 
\ .
\eea
The next-to-leading order equations then reduce to
\bea
\ddot \Phi_1 + 4 \, {\root 8 \of 8}\, e^{-\smallfrac{1}{4}\varphi}\,
\dot \Phi_1 + \smallfrac{3}{2} C e^\varphi = 0\ , & &
\\
\ddot \lambda_1 + {\root 8 \of 8}\, e^{-\smallfrac{1}{4}\varphi}\,
\left( - 2 \dot\Phi_1 + 8 \dot \lambda_1 + 5 \dot \nu_1 \right) -
8\, {\root 4 \of 8}\,  e^{-\smallfrac{1}{2}\varphi}\, \left( \Phi_1 - 5
\nu_1 \right) + \smallfrac{1}{4} C e^{\varphi} = 0 \ , & & 
\\
\ddot \nu _1 + 4\, {\root 8 \of 8}\,  e^{-\smallfrac{1}{4}\varphi}\,
\dot \nu_1 + 8\, {\root 8\of 8}\,  e^{-\smallfrac{1}{2}\varphi} (
\Phi_1 - 5 \nu_1 ) + \smallfrac{1}{4} C e^\varphi = 0 \ , & & 
\eea
and are solved by
\bea
& & \Phi _1 = 
- \smallfrac{3}{8\,{\root 8\of 8}}  C e ^
{\smallfrac{5}{4} \varphi} \tau \ ,
\\
& & \lambda _1 = \smallfrac{3}{64} C e ^ \varphi \tau ^2 
-\smallfrac{15}{256\, {\root 8 \of 8}} C e ^ {\smallfrac{5}{4}
  \varphi} \tau \ ,
\\
& & \nu _1 = - \smallfrac{3}{32\, {\root 8\of 8}} C
e^{\smallfrac{5}{4} \varphi} \tau - \smallfrac{1}{32}
8^{-\smallfrac{5}{4}}C e^{\smallfrac{3}{2}\varphi} \ .
\eea
In principle one should also consider the higher order iterations. 
However, we shall stop here since
in the low energy action we are already neglecting contributions of
higher order in $1/N$, as for example a cosmological constant.

Collecting the various contributions, one finds the following expressions
for the dilaton 
\beq
\Phi = \varphi - \smallfrac{3}{8} \, \smallfrac{1}{{\root 8\of 8}}
\, {C\over N} \, 
e^{{5\over 4}\varphi} \, \tau \ ,
\eeq
and for the metric tensor
\bea
{1\over \alpha '} ds^2  &=& 
d\tau ^2 + 
 \exp \left[ 2\, {\root 8\of 8}\, 
e^{-\smallfrac{1}{4}\varphi} \, \tau + \smallfrac{3}{32} \, \frac{C}{N}
e^\varphi \, \tau^2 \right] \eta_{ab} dx^a dx^b 
\nonumber 
\\ 
& & + \exp \left[ -\smallfrac{1}{4} \log 8 + \smallfrac{1}{2}\,
\varphi - \smallfrac{3}{16\, {\root 8 \of 8}} \, \frac{C}{N}\,
e^{\smallfrac{5}{4} \varphi} \, \tau \right]
d\Omega_5 ^2 \ , \label{metric-tau}
\eea

It is not hard to realize that the introduction of a dilaton tadpole
results into a running dilaton and running radii for the $AdS_5$
and $\bb{RP}^5$ spaces. This can be better appreciated if one parametrizes
the metric as
\beq
{1\over \alpha '} ds^2   =  R^2(u)\, {du^2 \over u^2 } + 
{u^2 \over R^2(u)} \, \eta_{ab}
dx^a dx^b +  \tilde R ^2(u) \, d\Omega _5 ^2 \ ,
\label{metric-u}
\eeq
in terms of the energy scale
\beq
u = \int e^{\lambda(\tau)} \, d\tau  \ ,
\eeq
with the radii 
\beq
R(u) = {d\tau \over d\log u}\ , \qquad \tilde R (u) = e^{\nu (\tau
(u))} \ . 
\eeq
Collecting our previous results, we find
\bea
\tau &=& \smallfrac{1}{{\root 8\of 8}}\, 
e^{\smallfrac{1}{4}\varphi}\, \left( \log u 
- \smallfrac{3}{64} \, \smallfrac{C}{N}\, e^\varphi \, \log ^2 u
\right)\ ,
\label{energy}
\\
R^2(u) &=& { \smallfrac{1}{{\root 4 \of 8}} \, e^{\smallfrac{1}{2}\varphi}
 \over 1 + \smallfrac{3}{16} \, {C\over N}
 \, e^{ \varphi}\, \log u } \ , 
\label{radius} 
\\
\tilde R ^2(u) &=& { \smallfrac{1}{{\root 4 \of 8}} \, e
 ^{\smallfrac{1}{2}\varphi} \over 1 +
\smallfrac{3}{16 \, {\root 4 \of 8}}\,  \smallfrac{C}{N}\,
 e^{2\varphi} \, \log u } \ ,
\label{radius-til} 
\\
e^{-\Phi} &=& e^{-\varphi} + \smallfrac{3}{8\, {\root 4 \of 8}}
\, \smallfrac{C}{N}\, e^\varphi \, \log u \ .
\label{dilaton}
\eea

At this point, before we interpret these results, various
comments are in order. First, in the above expressions
we are repeatedly ignoring  $1/N^2$ sub-leading terms.
Moreover, the value of $e^ \varphi$ can not be
interpreted any longer as the 't Hooft coupling, since now
the field theory is not conformal and the gauge coupling runs. 
In addition, from eqs. (\ref{radius}) and
(\ref{radius-til}) one might argue that $R$ and $\tilde R$
develop singularities at some point in the IR (notice that
for our model $C$ is positive and then singularity presents itself 
for $\log u <0$).  This, however, is not the case. The singularities are at 
points beyond the validity of the $1/N$ approximation and are 
an artifact of our choice of coordinates. Note, in fact, that the metric 
\eqref{metric-tau} is well defined for every energy scale. 
Finally, contrary to the familiar ${\cal N} =4$ super-conformal case
the radii $R$ and $\tilde R$ have now different values.
This is expected since at order $1/N$ the dilaton tadpole
behaves like an effective cosmological constant, and therefore 
(at that order) the overall curvature of the ten-dimensional space is 
no longer vanishing.

We can now turn to the interpretation of eqs. \eqref{radius} and 
\eqref{dilaton}, in the spirit of AdS/CFT correspondence. 
The behavior of the dilaton is consistent with the fact that,
in the dual gauge theory, no longer conformal, the coupling 
constant is expected to run. More precisely, once the dilaton is
identified with the 't Hooft coupling, $e^{\Phi(u)} \sim \alpha (u)$, 
eq. \eqref{dilaton} reproduces precisely the
logarithmic behavior \eqref{1loop} expected from the gauge theory
side, as already shown in \cite{AA}. Moreover, in the large 
$N$ limit the running of the coupling constant is suppressed, in 
agreement with the fact that in this limit the theory is conformal.
Following \cite{AA}, it would be interesting to compute also in this
case the numerical value of the dilaton tadpole $C$ and compare it with
the one-loop beta-function coefficient $b_1 = {16\over 3}$. To this
end, however, one should compute the contributions from closed strings in the
$AdS_5 \times \bb{RP}^5$ background with unoriented world-sheet wrapping the
$\bb{RP}^5$. Unfortunately, computations in genuinely curved backgrounds are 
still an open problem in string perturbation theory.

Following \cite{mal} the running of the AdS radius plays a natural
role in the computation of the quark anti-quark potential,
the subject of the next section.

\section{The Wilson loop}

We can now use the ``running'' $AdS_5 \times \bb{RP}^5$ geometry 
to compute the quark anti-quark potential using the holographic 
Wilson loop (see \cite{cobi} for a recent review). 
In \cite{rey,mal}  it was suggested that a natural 
description of the Wilson loop in terms of gravity is
\beq
\langle W({\cal C} ) \rangle \sim e^{-S} \ ,
\eeq
where, in the limit of large 't Hooft coupling, $S$ is the proper area of
a string world-sheet 
\beq
S= {1\over 2\pi\alpha '} \int d\tau d\sigma \, \sqrt{ {\rm det}
\ g_{\mu\nu} \, \partial_\alpha X^\mu \partial_\beta X^\nu }
\eeq
describing the loop ${\cal C}$ on the boundary of AdS. The quark anti-quark
potential can then be extracted choosing the standard rectangular loop
with sides $L$ and $T$, with $T\to \infty$.

Using the parametrization \eqref{metric-u} for the metric $g_{\mu\nu}$,
the Nambu-Goto action reads
\beq
S/T = E = \int dx \sqrt {(\partial _x u)^2 + u^4/ R^4(u)} \ ,
\eeq 
and thus the quark anti-quark separation and the
interaction energy are \cite{mal}
\beq
L \sim {R_0^2 \over u_0} \int {dy \over y^2 \sqrt {y^4 -
R^4_0/ R^4(y u_0)}} \ ,
\eeq
and
\beq
E \sim u_0 \int {y^2 dy \over \sqrt {y^4 - R^4_0/ R^4(y u_0 )}} \ ,
\eeq
with $R_0 =R(u_0)$.

If the energy $u_0$ is within the range of validity of 
the expansion, ${1\over N} \log u_0 \ll 1$, 
the ratio $R^4_0/R^4(y u_0)$ can be approximated by 
\beq
{R^4_0 \over R^4(y u_0)} \sim {1 + \smallfrac{3}{8} \smallfrac{C}{N} \,
e^\varphi \, \log y u_0 \over 1 + \smallfrac{3}{8} \smallfrac{C}{N} \,
e^\varphi \, \log  u_0}  
\sim 1 + \smallfrac{3}{8} \smallfrac{C}{N}
 e^{\varphi}\log y \ .
\eeq
Thus, one has
\beq
L\sim {R^2_0 \over u_0}\ , \qquad {\rm and}\quad E \sim u_0 \ ,
\eeq
and the potential 
\beq
E \sim {R^2_0\over L} \sim {R^2 (1/L) \over L} = {{1\over {\root 4 \of 8}}\,
e^{\smallfrac{1}{2}\varphi} \over L (1  - \smallfrac{3}{16} {C\over N}
 \, e^{ \varphi} \log L )} 
\label{qqbpot}
\eeq
can be expressed in terms of the running AdS radius, similarly to
 \cite{minahan}.

Eq. \eqref{qqbpot} reproduces precisely the behavior
expected from the gauge theory side:
at finite $N$ the Coulomb phase is characterized by a quark anti-quark 
potential \eqref{asymptotic} with a (logarithmically) running gauge coupling. 
The fact that the two prescriptions, 
the dilaton running 
\beq
E = {\alpha (L) \over L} = {e^{\Phi (L)} \over L} \sim
{e^{\varphi} \over L \left(1 - \smallfrac{3}{8 {\root 4 \of
8}}\, {C\over N} \, e^{2\varphi} \, \log L \right)} \label{qqbperb}
\eeq
and the Wilson loop \eqref{qqbpot}, yield similar results (in
contrast to tachyonic type 0 models \cite{log,minahan}) is
encouraging, although different powers of $e^\varphi$ are still involved in
the two cases.
Note that \eqref{qqbperb} is a perturbative prescription in contrast
to \eqref{qqbpot} that should, a priori, be applicable to the
non-perturbative regime.

Finally, let us describe the IR behavior of the gauge theory.
To this end, it is simpler to work with the metric in the form
\eqref{metric-tau}. The quark anti-quark energy is then
\beq
E = \int dx  \sqrt {e^{\lambda (\tau)} {(\partial _x \tau )}^2 +
 e^{2\lambda (\tau)}} \label{qqlong} \ ,
\eeq 
and, since 
\beq
\lambda (\tau) = \alpha \tau + {1\over N} \beta \tau ^2 \ ,
\eeq
with $\alpha$ and $\beta$ positive constants, $e^\lambda$ has
a minimum at $\tau_{{\rm min}} = - \alpha N / 2 \beta$, that corresponds to the
IR, implying a confining long distance potential \cite{KSS}
\beq
E \sim e^ {\lambda _{{\rm min}}} L = \sigma L \ .
\eeq
The reason behind this result is rather simple to understand.
For large enough separation, the string
will prefer to stay at the minimum of $\lambda$, thus minimizing 
the action \eqref{qqlong}. Therefore the
energy will be simply proportional to the separation on the AdS
boundary. Moreover, the QCD string tension (measured in units of
$1/\alpha '$) \cite{KSS}
\beq
\sigma = \exp \lambda _{{\rm min}} 
= \exp \left( -{\alpha ^2 N \over 4\beta} \right) = \exp
\left( - \smallfrac{16}{3} 
{\root 4 \of 8}\, e^{-\smallfrac{3}{2} \varphi}\, C^{-1}\, N\right) 
\eeq 
is exponentially suppressed with $N$, in agreement with field 
theory analysis \eqref{stringtension}. 

\section{Comments}

In conclusion, let us make a few comments about the validity of 
the holographic description that we have just presented. 
The holographic conjecture
relates 4d {\it strongly coupled} gauge theories to 5d
(super)gravity, and is thus valid, in principle, only within a certain 
range of energies. Moreover, the procedure of solving the
gravitational equations by iterations further restricts the 
solution. First, we are taking into account only the next-to-leading
contribution associated to the dilaton tadpole (a $1/N$ effect), and
therefore we need $N \gg 1$ to be sure that higher order string loops
be negligible. Moreover, the procedure of solving the
differential equations by iterations is valid only when the expansion
parameter is small. Since the expansion parameter is 
$\epsilon \equiv {1\over N} e^{\smallfrac{3}{2}\varphi} \log u$, 
the solution is valid only in some regions of the $(u\,,\, N)$
space. Note, however, that for any given energy $u$ and 
initial AdS radius $e^{\smallfrac{1}{4}\varphi}$, it is always
possible to find an $N$ such that $\epsilon$ is small, compatibly
with the previous requirement. Finally, the curvature of the space
should be small enough to allow one to trust classical gravity and to neglect
$\alpha '$ corrections. Once the previous requirements are met, this
last constraint simply translates into large  
$e^{\smallfrac{1}{4}\varphi}$, as in the \Nfour case.

Let us now see whether our solution meets these requirements.
It is not hard to see that large $e^{\varphi}$ and small expansion 
parameter $\epsilon$ translate into large 't Hooft coupling. 
But, from the field theory considerations, one expects
the logarithmic behavior for small and intermediate coupling \eqref{1loop}.
One should assume non-large $e^{\varphi}$ and
therefore $\alpha '$ corrections should not be neglected,
thus invalidating our solution. However, this is not the case.
A more careful analysis reveals that our results depend 
only on the existence of an $AdS_5 \times \bb{RP}^5$ solution at infinite
$N$. Now, since it is a common lore that $\alpha '$ corrections do
not change the geometry of the space \cite{green} (namely, 
that \Nfour is described by an $AdS_5 \times S^5$ metric at any 
coupling), the logarithmic running should not be affected.
Moreover, there seem to be problems in the computation of the 
short-distance Wilson loop. First, the direct Wilson loop computation results
into a value for the one-loop beta-function different from that
obtained by the running 
dilaton recipe. This is actually related to $\alpha '$ corrections
that, as in the \Nfour case \cite{green}, are expected to affect 
the parameters, but not the qualitative picture. In addition, 
the string world-sheet ``invades'' also regions where $\epsilon$
is not small. Namely, the holographic computation of the Wilson loop
for a given quark anti-quark separation $L$ requires that the metric be
known everywhere throughout the range $(u\sim 1/L\,,\,\infty)$. 
In order to avoid this problem and thus be able to trust our solution, 
the world-sheet should extend
deeply inside AdS, and thus the  potential \eqref{qqbpot} applies only to 
not too small quark anti-quark separations ({\it i.e.} our results are
not valid in the extreme UV).
Finally, the confining phase for long distances relies on the
existence of a minimum for $\lambda(\tau)$. However, at $\tau =
\tau_{{\rm  min}}$ the contribution of next-to-leading term
$\lambda_1$ is of the same order as $\lambda _0$. Thus, we can not
exclude that higher order contributions could eliminate the minimum, but
hopefully the only effect of higher corrections could be simply to shift
$\tau_{{\rm min}}$. However, a definite answer to this problem
requires an {\it exact} solution to the {\it exact} low-energy action.

\vskip 24pt
\noindent
{\bf Acknowledgments.} We are grateful to E. Gardi,
A. Sagnotti, D. Seminara and 
 especially to G. Grunberg and J. Sonnenschein for useful comments.
A.A. thanks the Department of Applied Mathematics and Theoretical
Physics at Cambridge University 
for the warm hospitality while this work was being completed. 
This research was supported in part by EEC under TMR contract 
ERBFMRX-CT96-0090.


\begin{thebibliography}{99}

\bibitem{ads} J. M. Maldacena,  Adv. Theor. Math. Phys. 2 (1998) 231;
S. S. Gubser, I. R. Klebanov and A. M. Polyakov, Phys. Lett. B428
(1998) 105;
E. Witten,  Adv. Theor. Math. Phys. 2 (1998) 253. See also 
O. Aharony, S. S. Gubser, J. Maldacena, H. Ooguri and
 Y. Oz, hep-th/9905111 for a recent review.

\bibitem{KT1} I.R. Klebanov and A.A. Tseytlin,
Nucl. Phys. B546 (1999) 155; JHEP 9903 (1999) 015;
M. Bill\'o, B. Craps and F. Roose, Phys. Lett. B457 (1999) 61; 
A. Armoni and B. Kol, JHEP 9907 (1999) 011.

\bibitem{BFL2} R. Blumenhagen, A. Font and D. L\"ust, Nucl. Phys. B560
(1999) 66.

\bibitem{minahan} J. A. Minahan, JHEP 9904 (1999) 007.

\bibitem{log}
 J.A. Minahan, JHEP 9901 (1999) 020;
 I.R. Klebanov and A.A. Tseytlin, Nucl. Phys. B547 (1999) 143;
A. Armoni, E. Fuchs and J. Sonnenschein, JHEP 9906 (1999) 027;
M. Alishahiha, A. Brandhuber and Y. Oz, JHEP 9905 (1999) 024.

\bibitem{klebanov} I. R. Klebanov, Phys. Lett. B466 (1999) 166.

\bibitem{AA} C. Angelantonj and A. Armoni,  hep-th/9912257, to appear
in Nucl. Phys. B.

\bibitem{sagnotti2} A. Sagnotti, hep-th/9509080, hep-th/9702093.

\bibitem{KS} S. Kachru and E. Silverstein, Phys. Rev. Lett. 80 (1998) 4855;
Z. Kakushadze, Nucl. Phys. B529 (1998) 157.

\bibitem{ADS} C. Angelantonj, Phys. Lett. B444 (1998) 309;
R. Blumenhagen and A. Kumar, Phys. Lett. B464 (1999) 46;
S. Sugimoto, Prog.Theor.Phys. 102 (1999) 685;
I. Antoniadis, E. Dudas and A. Sagnotti, Phys. Lett. B464 (1999) 38; 
G. Aldazabal and A. M. Uranga, JHEP 9910 (1999) 024;
K. F{\"o}rger, Phys. Lett. B469 (1999) 113;
C. Angelantonj, I. Antoniadis, G. D'Appollonio,
E. Dudas and A. Sagnotti, hep-th/9911081;
M. Bianchi, J.F. Morales and G. Pradisi, hep-th/9910228; 
A. M. Uranga, hep-th/9912145.

\bibitem{BFL1} R. Blumenhagen, A. Font and D. L\"ust, Nucl. Phys. B558
(1999) 159.

\bibitem{PC} J. Polchinski and Y.C. Cai, Nucl. Phys. B296 (1988) 91;
M. Bianchi and F. Morales, hep-th/0002149.

\bibitem{KNT} I. R. Klebanov and N. A. Nekrasov, hep-th/9911096;
 I. R. Klebanov and A. A. Tseytlin, hep-th/0002159.

\bibitem{AlGo} 
E. Alvarez and C. Gomez, Nucl.Phys. B566 (2000) 363-372;
J. de Boer, E. Verlinde and H. Verlinde, hep-th/9912012;
E. Verlinde and H. Verlinde, hep-th/9912018;
E. Alvarez and C. Gomez, hep-th/0001016.

\bibitem{witten} E. Witten, JHEP 07 (1998) 006.

\bibitem{pst} P. Pasti, D. Sorokin and M. Tonin, Phys. Rev. D55 (1997) 6292.

\bibitem{cobi} J. Sonnenschein, hep-th/0003032.

\bibitem{mal} J. M. Maldacena,  Phys. Rev. Lett. 80 (1998) 4859.

\bibitem{rey} S.-J. Rey and J. Yee, hep-th/9803001.

\bibitem{KSS} Y. Kinar, E. Schreiber and J. Sonnenschein,
Nucl. Phys. B566 (2000) 103.

\bibitem{green} T. Banks and M.B. Green, JHEP 9805 (1998) 002;
R. Kallosh, A. Rajaraman, Phys.Rev. D58 (1998) 125003.

\end{thebibliography}
\end{document}